# An Explorative Study of GitHub Repositories of AI Papers

Boyang Zhang

*Key Laboratory of High Confidence Software Technologies (Peking University), Ministry of Education, PRC*

boyangzhang@pku.edu.cn

**Abstract**

With the rapid development of AI technologies, thousands of AI papers are being published each year. Many of these papers have released sample code to facilitate follow-up researchers. This paper presents an explorative study of over 1700 code repositories of AI papers hosted on GitHub. We find that these repositories are often poorly written, lack of documents, lack of maintenance, and hard to configure the underlying runtime environment. Thus, many code repositories become inactive and abandoned. Such a situation makes follow-up researchers hard to reproduce the results or do further research. In addition, these hard-to-reuse code makes a gap between academia and industry. Based on the findings, we give some recommendations on how to improve the quality of code repositories of AI papers.

## I. Introduction

With the rapid development of AI technologies, thousands of AI papers are being published each year. Many of these papers have released sample code, which can greatly facilitate follow-up researchers. However, these code has some problems. They often lack of documents, lack of maintenance, and hard to configure the underlying runtime environment. Such problems make these code hard to read, run, and modify. Many of these repositories become inactive, abandoned, and forgotten soon after they were published. These problems make a barrier to follow-up researchers to reproduce the results or do further research. Also, these code cannot be reused in a modular way. As a result, the code is hard to interoperate with other software, making a gap between academia and industry.

In this paper, we present an explorative study on over 1700 code repositories on GitHub. We concluded several patterns of best and worst repositories. Based on the findings, we give several recommendations on how to improve the quality of code repositories of AI papers. Future researchers can make their code more useful based on our findings.

## II. Dataset

Sample code is often published on GitHub. Volunteers have organized a list of papers with sample code attached [1]. This list is actively maintained. Therefore we choose this list as our dataset. The list is organized in CSV format. Each record of this list contains the link to a paper, the conference name where the paper was published, the link to the corresponding code repository, and the star count of the code repository.

We crawled all repositories in that list. Due to some invalid links, the total number of repositories is slightly smaller than the total record count. The reason why some links are invalid is that the repositories may be deleted or made private by their author. We watched these links for one month.

During this period, we observed serval repository disappearance. Finally, we cloned 1798 repositories in total from GitHub.

## III. Findings

In this section, we first analyze the repositories from three aspects, i.e., documentation, maintenance and reusability. Then we manually set up some top repositories to examine the real efforts needed to reuse the code.

### 1. Documentation

Popular open source projects often have dedicated websites, easy-to-read documents, high-quality code comments, and active community which consists of many contributors. In contrast, sample code from AI papers is usually written by individuals. The code are often self-explanatory, thus few code has high-quality comments. The documents mainly lied in readme files, instead of dedicated document files. In order to measure the quality of documents, we count number of lines of readme documents of each repository.

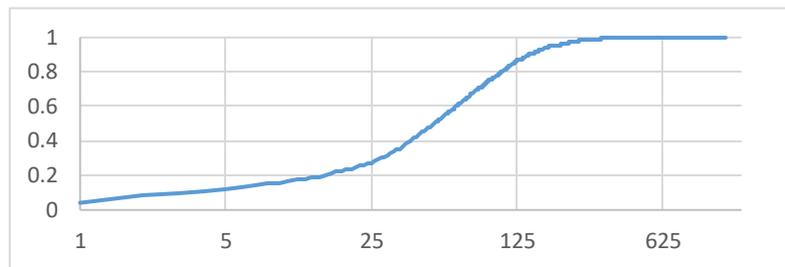

Figure 1 Distribution of line count of readme document

As we can see in Figure 1, 3% of the repositories have no readme document, 12% have less than 5 lines of readme document, and 30% have less than 30 lines of readme document. This result indicates that many authors provide little information about their repositories in the readme document.

### 2. Maintenance

In order to measure how repositories are maintained, we propose two indicators. Figure 2 shows the distribution of total commit count in single repository. Figure 3 shows the distribution of how many days has lasted after latest commit of one repository as of 2019/1/16. It's needed to point out that these two indicators does not take repository publication time into account, so it will make the statistics more optimistic.

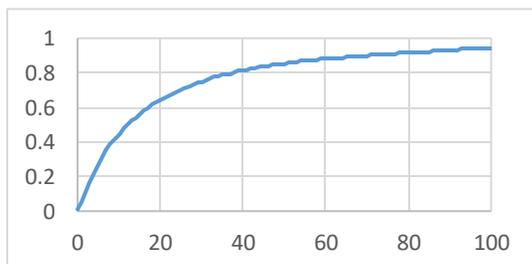 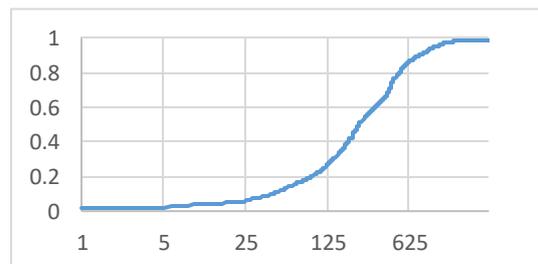

Figure 2 Distribution of total commit count

Figure 3 Distribution of last commit time
(how many days before 2019/1/16)

Figure 2 shows that 44% of them have less than 10 commits. Figure 3 shows that only 20% of them have commits in last 90 days. The result implies that many repositories are inactive. The reason might be the authors just publish the code and do not care about it anymore.

**3. Reusability**

In recent years, Docker [2] became an emerging way to deploy software. A docker image consists of application program and its underlying runtime environment. With pre-created docker images, a user can deploy an application easily and conveniently. *Dockerfile* contains the instruction to build a docker image. If a repository provides a *Dockerfile*, a user can build the docker image and start running the program in a few steps.

Anaconda [3] is a python distribution. It has a powerful tool named Conca to manage runtime environment of a code repository. The *environment.yml* contains the environment configuration of Conda. It can greatly simplify the procedure of configuring the runtime environment.

If a repository contains one of *Dockerfile* or *environment.yml*, it is likely that a user can setup the runtime environments and start experimenting in a few steps. However, very few of repositories provide these files. The statistics are listed in Table 1.

Table 1  Number of repos containing at least one *Dockerfile* or *environment.yml*

| Files | # of repos contains at least one … file | Percentage |
|---|---|---|
| *Dockerfile* | 73 | 4.1% |
| *environment.yml* | 19 | 1.1% |
| Total | 92 | 5.1% |

This result means that although there are ways to help user configure the runtime environment conveniently, most authors do not provide a configuration file. The reason may be that the authors are unfamiliar with Docker or Conda, or they are just too lazy to provide them.

**4. Lessons from top repositories**

In order to find out the real difficulties for researchers to reuse the repositories, we chose some most-stared repositories which were published in 2018, then try to configure them manually and run their demo experiments. We analyzed the installation procedure and encountered difficulties.

Table 2 Analysis of the procedure to reproduce experiment results

| Title or repo name | Stars | Rank | Features and problems |
|---|---|---|---|
| Video-to-Video Synthesis | 5578 | #1 | Dockerfile provided, well-written documents, easy to install, can reproduce expected results. |
| Deep Image Prior | 3736 | #2 | Dockerfile provided, easy to install, interactive jupyter notebooks, can reproduce expected results. |
| PRNet | 2434 | #4 | Well-written documents, need install dependencies manually, can reproduce expected results. |
| Learning-to-See-in-the-Dark | 2326 | #5 | Well-written documents; however dataset can't be downloaded because of Google Drive policies, we are unable to run it[1] |
| glow | 2088 | #6 | Well-written documents, need to install dependencies manually; however the dependencies is complicated, we are unable to setup it in few hours |

---

[1] The dataset download link became valid few days after our experiment.

We concluded the main reasons why these repositories are so popular. First, they have well-written documents to indicate users how to reproduce their impressive experiment results. Second, the less steps required to setup environment, the easier to reproduce result successfully. Finally, the author can respond to user's feedbacks and give them help when they encountered problems.

However, these top repositories also have problems. Sample code often requires huge dataset to be downloaded separately, increasing the possibility of failure because of downloading problems[2]. Besides, if the dependencies are difficult to setup, the failure rate will be higher.

We recommend future researchers use Docker or Conda as their dependency manager. These methods are much more stable than manually installing dependencies. It can also save a lot time for follow-up researchers.

## IV. Related Work

The increasing number of open source repositories make developing new software easier. However, it is hard to write high-quality open source projects. Zhu et al. [4] found the relation between popularity and patterns of folder use. Our work analyzed documentation quality, maintenance quality and code reusability of a few popular AI related repositories. Ray et al. [5] found the relation between code quality and programming language. In contrast, top AI related repositories often use Python as their programming language. Kalliamvakou et al. [6] found majority of GitHub projects are personal and inactive. Our paper also focused on the quality of maintenance of repositories. Jarczyk et al. [7] presented metrics for project's popularity and the quality of support offered by team members to users. We also pointed out the importance of the author's feedback.

## V. Conclusion

In this paper, we analyzed over 1700 code repositories released by AI papers. We find these repositories are often poorly written, lack of documents, lack of maintenance, and hard to configure the underlying runtime environment. We also analyzed a few popular code repositories, and concluded three recommendations: well-written document, easy to setup, and responsive communication between authors and users. Our findings can help future researchers make their code more useful to follow-up researchers and industry. As for the future work, we plan to abstract each AI repository as a service and leverage the technique of service computing [8], [9], [10], [11] to facilitate the reusability of AI-related code.

---

[2] For example, Google Drive changes API once a few years. At that time, download scripts in these repositories will become invalid.